# A Case Study: Task Scheduling Methodologies for High Speed Computing Systems

[1]Mahendra Vucha and [2]Arvind Rajawat

[1,2]Department of Electronics & Communication Engineering,
Maulana Azad National Institute of Technology, Bhopal, India

## ABSTRACT

*High Speed computing meets ever increasing real-time computational demands through the leveraging of flexibility and parallelism. The flexibility is achieved when computing platform designed with heterogeneous resources to support multifarious tasks of an application where as task scheduling brings parallel processing. The efficient task scheduling is critical to obtain optimized performance in heterogeneous computing Systems (HCS). In this paper, we brought a review of various application scheduling models which provide parallelism for homogeneous and heterogeneous computing systems. In this paper, we made a review of various scheduling methodologies targeted to high speed computing systems and also prepared summary chart. The comparative study of scheduling methodologies for high speed computing systems has been carried out based on the attributes of platform & application as well. The attributes are execution time, nature of task, task handling capability, type of host & computing platform. Finally a summary chart has been prepared and it demonstrates that the need of developing scheduling methodologies for Heterogeneous Reconfigurable Computing Systems (HRCS) which is an emerging high speed computing platform for real time applications.*

## 1.INTRODUCTION

Microprocessor is at the core of high performance computing systems but they provide flexible computing at the expense of performance [1]. Application Specific Integrated Circuit (ASIC) supports fixed functionality and superior performance for an application but they restrict flexibility of architecture. Thereafter a new computing paradigm [2] Reconfigurable Systems (RS) promises greater flexibility without compromise in performance. So complex applications like MIMO, OFDM and image processing are accelerated by reconfigurable architecture and achieved higher performance by reducing the instruction fetch, decode and execute bottleneck [1][2][3]. The RS brings the phenomenon of configuring custom digital circuits dynamically and modified via software. This ability of creating and modifying digital logic circuits without physically altering the hardware provides more flexible and low cost solution for real time applications. This phenomenon of dynamic reconfiguration of an application is enabled by the availability of high density programmable logic chips called Field Programmable Gate Array (FPGA). So, High Speed Computing Systems (HSCS) should have one or more resources of such kind (Reconfigurable System on Chip (RSoC) [15], MOLEN architecture [26]) as Processing Element (PE) to enhance the speed of real time application. A computing platform described in [26][27][28][29] and these are made by integrated similar resources through high speed network to support the execution of parallel applications called *Homogeneous Computing System*. The efficiency of homogeneous computing system critically depends on the methods used in [22][23][24][26] to schedule tasks of parallel applications. Other hand, diverse set of resources interconnected with a high speed networks provides a new computing platform [20][21] called *Heterogeneous Computing System*, which could support executing computationally intensive parallel and distributed applications. An emerging computing platform which integrates the array of programmable logic resources and soft core processors together on a single chip [15][16][17][18] called *Heterogeneous Reconfigurable Computing Systems* (HRCS). The HRCS platform is an emerging paradigm of research that offers cost effective solutions for computationally intensive applications through hardware reuse and many multimedia applications [2] were accelerated by HRCS.

In real time, tasks of parallel application must share the resources of HSCS effectively in order to enhance the execution speed of an application and it could be achieved through effective scheduling mechanism. There are many researchers presented techniques for mapping multiple tasks to HSCS with the aim of "minimizing execution time of an application' and also "efficient utilization of resources". In this paper, we bring the review of various existing scheduling methodologies for HSCS. The task scheduling models are basically two types called static and dynamic scheduling. *Static Scheduling:* All information needed for scheduling such as the structure of the parallel application, execution time of individual tasks and communication cost between the tasks must be known in advance [10][12][13][14]. *Dynamic scheduling:* The scheduling decisions made at runtime and whereas its aim is not only enhance the execution time and also minimize the communication overheads [8][20][24][26]. The review of static and dynamic scheduling heuristics for HSCS is described thoroughly in next chapters. In general, various scheduling heuristic approaches are classified into four categories: List scheduling algorithms [20], clustering algorithms [11], Duplication Algorithms [22], and genetic algorithms. Among them, the list scheduling algorithms provides good quality of schedule and their performance is compatible with all categories of applications [20]. So in this paper, we have concentrated more on list scheduling algorithm and it has three steps: task selection, processor selection and status update. For clear understanding, the remaining paper is organized as the task scheduling for homogeneous computing systems in chapter 2, heterogeneous computing systems in chapter 3, reconfigurable computing systems in chapter 4, heterogeneous reconfigurable computing systems in chapter 5, and review summary chart in chapter 6 and finally paper is concluded in chapter 7.

## 2. SCHEDULING MODELS FOR HOMOGENEOUS COMPUTING SYSTEMS

Homogeneous Computing refers the systems which are formulated with multiple similar kinds of soft core processors and it brings parallelism for application execution. The parallelism is achieved by effective task scheduling. The static and dynamic list scheduling techniques are summarized [5] for microprocessor based systems. In [5], the task scheduling is based on cost function whereas the cost function is an attribute of tasks of an application. There are several list scheduling algorithms proposed [5][27][28][29] for microprocessor as follows. *Rate Monotonic Algorithm:* The Rate Monotonic (RM) algorithm [27] is a static priority based scheduling algorithm, which assigns the highest priority to the most frequency task and lowest priority to least frequency task in the system. The RM selects the highest priority task to execute first and then remaining tasks come for execution as per their priority sequence. So the RM can only used in statically defined systems and the scheduling bound of RM algorithm is less than 100%. So that the researchers in task scheduling moved towards the use of dynamic priority based scheduling algorithms. *Earliest Deadline First Algorithm:* The Earliest Deadline First (EDF) algorithm [28] uses the dead line of the task as cost function. The task with earliest deadline has highest priority whereas the task with longest deadline has lowest priority. The major advantages of EDF algorithm is that the priorities are dynamic so the period of tasks can be changed dynamically and also the schedulable bound for any task set is 100% but there is no control of which tasks fails during transient overload. *Minimum Laxity First Algorithm:* The Minimum Laxity First (MLF) algorithm [29][4] follows dynamic scheduling where it assigns Laxity to each task in a system and selects the task having minimum laxity to execute next. The Laxity is a measure of flexibility of a task to schedule and it is defined as follows: Laxity = deadline time – current time – executing time. It also has 100% schedulable bond like EDF and there is no way to control which tasks are guaranteed to execute during a transient overload. *Maximum Urgency First Algorithm:* The Maximum Urgency First (MUF) algorithm [29] follows both static and dynamic priority scheduling. In MUF algorithm, each task would be given with an Urgency and the Urgency is combination of two fixed priorities and one dynamic priority. The static priorities are defined once and do not changed during execution where as the dynamic priority is assigned at runtime which is inversely proportional to the Laxity of a task. The MUF scheduler looks first in static priority and then dynamic priority. A low cost task scheduling [26] described for Distributed Memory Machines (DMM) based on the heuristics EDF, MLF etc. and stated that the List Scheduling with Dynamic Priorities (LSDP) gives optimum results than List Scheduling with Static Priority (LSSP). The task duplication based scheduling [22] for distributed memory machines designed to reduce the inter processor communication. A Modified TDS (MTDS) described in [23] and it generates shorter scheduled list then TDS [22]. A Dynamic Critical Path (DCP) Scheduling algorithm [24] proposed for multiprocessors where the DCP intended to find critical path of a task graph and rearranges the schedule on each processor dynamically. A duplication-based scheduling strategy called Selective Duplication (SD) algorithm is developed [25] for multiprocessor systems with the aim of exploit the available scheduling holes effectively without scarifying efficiency. In [25], the application is visualized as DAG and the targeted machine is represented as $M = (P, [L_{ij}], [h_{ij}])$; P = {p1, p2, ...,Pp} is set of P homogeneous processor; $[L_{ij}]$ is a $p \times p$ matrix describing interconnection network topology and $[h_{ij}]$ is a $p \times p$ matrix giving minimum distance in number of hops between processor $p_i$ and $p_j$. The SD algorithm [23] is compared with existing duplication TDS [22], MTDS [23], and non-duplication scheduling algorithms with respect to Normalized Schedule Length (NSL), Efficiency.

## 3. SCHEDULING MODELS FOR HETEROGENEOUS COMPUTING SYSTEMS

Heterogeneous computing refers to systems that have more than one kind of processing elements and it gains performance for the application when multifarious execution required. An application scheduling algorithms called *Heterogeneous Earliest Finish Time* (HEFT) and *Critical-Path-On-a-Processor* (CPOP) formulated [20] for a bounded number of Heterogeneous processors. The HEFT has two phases, Task prioritizing phase uses HEFT as cost function and processor selection phase to select the tasks on its best processor. The HEFT has the time complexity $O(e \times q)$ for e edges and q processors. The CPOP used Critical Path, which is sum of computation time and inter task communication time, as cost function and provides time complexity equal to $O(e \times p)$ for *e* edges and *p* processors. The HEFT algorithm outperforms other algorithms in terms of SLR and Speedup but the CPOP algorithm outperforms the related work in terms of average SLR. On an average, the HEFT [20] algorithm is faster than the CPOP algorithm by 10 percent, the Mapping Heuristic (MH) algorithm by 32 percent, the *Dynamic Level Scheduling* (DLS) algorithm by 84 percent, Levelized-Min Time (LMT) algorithms by 48 percent. A high performance static scheduling algorithm [21] called Longest Dynamic Critical Path (LDCP) algorithm presented for Heterogeneous Distributed Computing Systems (HeDCS). In order to compute the LDCP, the HeDCS is formulated with m heterogeneous processors and application is computed as Direct Acyclic Graph that corresponds to a Processor P$_j$ (DAGP$_j$) with size of task set to their computation cost on that processor $P_j$. The DAGP nodes are assigned with *upward rank [21]* (URank) and URank acts as cost function to prioritize them for scheduling whereas the Urank is summation of execution time on processor and communication cost between the adjacent tasks. The LDCP scheduling algorithm outperforms the both HEFT [20] and DLS algorithms in terms of Normalized Schedule Length (NSL) and speedup. A generalized fixed priority CPU scheduling model with the notion of pre-emption threshold [10] is developed and it bridges the gap between pre-emptive and non-preemptive scheduling models in real time. The scheduling model [10] addresses the problem of finding an optimal priority ordering and pre-emption threshold assignment for the tasks which are independent and do not suspend themselves whereas the overheads due to context switching are negligible. The model [10] introduces pre-emptablity as it is enough to achieve feasibility and ensures optimum schedulability by reducing scheduling overheads through minimum number of pre-emptions.

## 4. SCHEDULING MODELS FOR RECONFIGURABLE COMPUTING SYSTEMS

Reconfigurable Computing is an emerging paradigm that satisfies simultaneous demand for application flexibility and performance. The ability of customize its architecture, to support the concurrent computation and parallel application execution, demonstrates RCS performance benefits over the general purpose processor. A Parameterized Module Scheduling (PMS) algorithm for RCS [8][4]addressed the problem of scheduling and mapping non-preemptive tasks of an application task graph to platform having variable Reconfigurable Logic Units (RLUs) by the concept parameterized modules and variable silicon area. The scheduling system [8] follows the concept Dynamic Programming (DP) to schedule the tasks & it is described in three parts: application in the form of task graph, computing environment and performance criteria to obtain the scheduling goal. Here, performance criteria would be the scheduling length 'L' i.e. actual Finish Time (FT) of the exit task $v_{exit}$ (L = FT ($v_{exit}$)) and the goal is to minimize the scheduling length 'L' of an application. The scheduling algorithm [8] uses the b-level of task as rank function to prioritize the tasks of an application where b-level of a task node $V_i$ is the length of longest path from the node $V_i$ to exit task node. Loop Kernel Pipelining Mapping (LKPM) [2] addressed for Coarse Grained Reconfigurable Architecture (CGRA) to optimize Data Intensive Applications (DIA). In [2], The Program Information Aided Control Dataflow Task Graph (PIA-CDTG) represents the functionality and behaviour of DIA, Virtual Instruction Dataflow Graph (Vi-DFG) represents the behaviour of critical loop kernels and Reconfigurable Architecture Graph (RAG) represents the loop self pipelining and loop iteration behaviour of CGRA. The M×N CGRA can be represented by RAG = (PE, C) where $PE_{ij}$ consists of memory PE (mPE) and computation PE (cPE), $0 \leq i \leq M, 0 \leq j \leq N$ and C describes the data relevance dependency. The LKPM map the control conditions of loop to mPEs and body of the loop to cPEs to increase the throughput of DIA. A dynamic scheduling and placement algorithm [11] has been proposed for RS based on finishing time mobility of the tasks. The model in [11] integrates an online placement algorithm with scheduling model to support FPGA clusters. Here the FPGA is divided into slots or clusters and the arriving tasks are placed inside one of the cluster depending on their execution end time values. To enhance the efficiency of the device [11], the width of the clusters varies in runtime when needed and the host processor could control the mapping of hardware task code as an executable circuit to FPGA. Online scheduling of real time tasks to reconfigurable computing systems [12][13][14] formalized with the objective of reducing configuration overheads through resource reuse and minimizes the total execution time in addition to decrease task rejection ratio. The model in [12] is combination of window based stuffing algorithm and KAMER [11] placement algorithm. The model in [13] focuses on real time independent tasks and the tasks are defined with 5 – tuple $T_i = \{w_i, h_i, e_i, a_i, d_i, r_i\}$ where $w_i, h_i, e_i, a_i, d_i \text{ and } r_i$ represents width, height, execution time, arrival time, dead line and reconfiguration time of tasks respectively. The schedulable bound of these algorithms [12, 13] is less than 100%. A heuristic approach to schedule periodic real time tasks on RH [14] formalized two scheduling algorithm called EDF-Next Fit (EDF-NF) and Merge Server Distribute Load (MSDL) for preemptive periodic tasks. The MSDL constructs a set of servers by properly merging set of tasks for parallel execution and the resulted servers are then scheduled for sequential execution on FPGA with EDF-NF.

## 5. SCHEDULING ALGORITHMS FOR HETEROGENEOUS RECONFIGURABLE COMPUTING SYSTEMS

A computing platform called MOLEN Polymorphic processor [26] presented and it is incorporated with both general purpose and custom computing processing elements. The MOLEN processor is also incorporated with arbitrary number of programmable units to support both hardware and software tasks. An efficient multi task scheduler for runtime reconfigurable systems [9] proposed a new parameter called Time-Improvement as cost function for compiler assisted scheduling algorithm. The Time-Improvement heuristic is defined based on reduction-in-task-execution time and distance-to-next-call. The scheduling system in [9] target to MOLEN Polymorphic processor [26] and it assigns less CPU intensive tasks and control of tasks to General Purpose Processor (GPP) whereas computing intensive tasks are assigned to FPGA. The task scheduler in [9] outperforms previous algorithms and accelerates task execution from 4% up to 20%. Online scheduling of Software Tasks (ST), Hardware Tasks (HT) and Hybrid Tasks (HST) proposed [6] for CPU-FPGA platform, where ST executes only on CPU, HT executes only on FPGA and the HST execute on both CPU & FPGA. The scheduling model [6] uses reserved time of tasks as cost function and it is integration of task allocation, placement and task migration modules. An On-line HW/SW partitioning and co-scheduling algorithm [3] proposed for GPP and Reconfigurable Processing Unit (RPU) environment in which Hardware Earliest Finish time (HEFT) and Software Earliest Finish time (SEFT) are calculated for tasks of an application. The difference between HEFT and SEFT imply to partition tasks and EFT used to define task scheduled list for GPP and RPU as well. An overview of Tasks co-scheduling is described [7][31] to µP and FPGA environment from different communities like Embedded Computing (EC), Heterogeneous Computing (HC) and Reconfigurable Hardware (RH). The Reconfigurable Computing Co-scheduler (ReCoS) [7] integrates the strengths of HC and RH scheduling to handle the RC system constraints such as the number of FFs, LUTs, Multiplexers, CLBs, communication overheads, reconfiguration overheads, throughputs and power constraints. The ReCoS algorithm as compared with EC, RC and RH scheduling algorithms, shows improvement in optimal schedule search time and execution time of an application. Hardware supported task scheduling for Dynamically RSoC [15] described to effectively utilize the RSOC resources for multi task applications. Task systems in [15] represented as modified Directed Acyclic Graph (DAG) and the task graph is defined as tuple $G = (V, E^d, E^c, P)$, where V is set of nodes, $E^d$ and $E^c$ are the set of directed data edges and control edges respectively and P represents the set of probabilities associated with $E^c$. The RSoC architecture in [15] comprises a general purpose embedded processor along with two L1 data and instruction cache and a number of reconfigurable logic units on a single chip. The summary of the paper [15] states that Dynamic Scheduling (DS) does not degrade as the complexity of the problem increase whereas the performance of Static Scheduling (SS) decline and finally the DS outperforms the SS when both task system complexity and degree of dynamism increases. Compiler assisted runtime scheduler [16] is designed for MOLEN architecture where the compiler describes the run time system as Configuration Call Graph (CCG). The CCG in [16] demonstrates two parameters called *the distance to the next call* and *frequency of calls in future* to the tasks and these parameters acts as cost function to the scheduler. Communication aware online task scheduling for partially reconfigurable systems [17] distributes the tasks to 2D area based on data communication time of tasks. The scheduler in [17] can run on host processor and tasks expected end time $t_c = t_{latest} + t_{config} + t_{comm1} + t_{exe} + t_{comm2}$, where $t_{latest}$ is completion time of already scheduled task $t_{config}$ is task configuration time, $t_{comm1}$ is data/memory read time, $t_{exe}$ is task execution time and $t_{comm2}$ is data/memory write time. HW/SW co-design techniques [18] are described for dynamically reconfigurable architectures with the aim of deciding execution order of the event at run time based on their EDF. Here authors have demonstrated a HW/SW partitioning algorithm, a co-design methodology with dynamic scheduling for discrete event systems and a dynamic reconfigurable computing multi-context scheduling algorithm. These three co-design techniques [18] minimizes the application execution time by paralleling events execution and controlled by host processor for both shared memory and local memory based Dynamic Reconfigurable Logic (DRL) architectures. When number of DRL cells is equal or more than three, the techniques in [18] brings better optimization for shared memory architecture than the local memory architectures. A HW/SW partitioning algorithm [30] presented to partition the tasks as software tasks and hardware tasks based on their waiting time. A layer model [20] provides systematic use of dynamically reconfigurable hardware and also reduces the error-proneness of the system components. The Layer Model [20] comprises of six layers (Bottom to Top). The lowest or first layer *Hardware Layer* represents the reconfigurable hardware, second Configuration *Layer* interfaces with the configuration port of FPGA, third *Positioning Layer* assigns the position to partial bit-stream, fourth Allocation *Layer* manages the resources for incoming modules on FPGA, fifth *Module Management Layer* provides access to all modules (tasks) that are loaded to the system and sixth *Application Layer* represent the application as task graph. These kind of layer models helps to design efficient operating for HRCS.

## 6. SUMMARY CHART FOR SCHEDULING MODELS FOR HIGH SPEED COMPUTING SYSTEMS

The summary chart of scheduling methodologies shown in table 1 demonstrates the author and paper reference in first column, nature of scheduling algorithm (static or dynamic) in second column, tasks handling behaviour (single task or multiple task supported) in third column, nature of computing resources in targeted computing platform ( microprocessor or FPGA or integration of µP and FPGA) in fourth column, nature of host platform where the scheduling methodology executes ( µP or FPGA) in fifth column, the targeted performance metrics (schedulable bound, execution speed enhancement and resources optimization) in sixth column, cost function, which is used to prioritize the tasks of an application that helps to prepare scheduled task list, in seventh column and finally future scope and remark of the methodology is described in eighth column. The overview and summary chart of various scheduling methodologies described for HSCS are as follows in table 1.

International Journal of Embedded systems and Applications(IJESA) Vol.4, No.4, December 2014Table 1: Summary chart of scheduling methodologies

| Approach \ Reference paper | Nature | | Task support | | Target Platform | | | Host Platform | | Performance Evaluation | | | COST function | Scope / Remark |
|---|---|---|---|---|---|---|---|---|---|---|---|---|---|---|
| | Static | Dynamic | Single | Multiple | Microprocessor (µP) | FPGA | µP + FPGA | µP | FPGA | Schedulable bound | Speed Enhancement | Resource Optimization | | |
| | | | | | | | | | | | | | | **RM** – Rate Monotonic<br>**EDF** – Earliest deadline First<br>**MLF** – Maximum Laxity First<br>**MUF** – Maximum Urgency First<br>**EFT** – Earliest Finish Time<br>**ALAP** – As Late As Possible<br>**AET** – Average Execution Time<br>**HEFT** - *Heterogeneous Earliest Finish Time*<br>*CPOP: Critical-Path-On-a- Processor* |
| [5] [27] | × | | × | | × | | | × | | × | | × | Frequency of the task | Schedulable bound is less than 100% |
| [5] [8] | | × | × | | × | | | × | | × | | | EDF | There is no control on which tasks fails during transient overload |
| [5] [29] | | × | × | | × | | | × | | × | | | MLF & MUF | MLF: There is no control on which tasks fails during transient overload<br>MUF: Combination of fixed and dynamic nature in order to avoid the drawback of MLF & EDF |
| . [26] | | × | | × | × | | | × | | × | | × | EDF | List Scheduling with Dynamic Priorities (LSDP) gives optimum results than List Scheduling with Static Priority (LSSP) |
| [22] [23] [25] | | × | | × | × | | | × | | | × | × | Selective Task Duplication | Designed for Distributed Memory Machines to reduce inter processor communication |
| [24] | | × | | × | × | | | × | | × | × | × | Dynamic Critical Path of task | It determines the critical path of the task graph and selects the next node to be scheduled in a dynamic fashion |
| [20] | | × | | × | × | | | × | | × | × | | HEFT & | HEFT algorithm is faster than the CPOP algorithm by 10 %, the |

| | | | | | | | | | | | | |
|---|---|---|---|---|---|---|---|---|---|---|---|---|
| | | | | | | | | | | | CPOP | Mapping Heuristic (MH) algorithm by 32%, the *Dynamic Level Scheduling* (DLS) algorithm by 84 %, Levelized-Min Time (LMT) algorithms by 48 %. |
| . [21] | × | | | × | × | | | × | | × | × | | upward rank | LDCP algorithm outperforms the both HEFT and DLS algorithms in terms of Normalized Schedule Length (NSL) and speedup. *(*Urank is summation of execution time on processor and communication cost between the adjacent tasks) |
| [10] | × | | | × | × | | | × | | × | × | | pre-emption threshold | It reduces scheduling overheads through minimum number of pre-emption. At most 18% improvement in execution |
| . [8] | | × | × | | | × | | × | | × | × | | Level of task graph | Minimization of Schedulable length |
| [2] | × | | × | | | × | | × | | | × | | Level of task graph & task dependencies | Program Information Aided Control Dataflow Task Graph (PIA-CDTG) represents the functionality and behaviour of Data Intensive Application and Virtual Instruction Dataflow Graph (Vi-DFG) represents the behaviour of critical loop kernels |
| [11] | | × | × | | × | × | | × | | | × | × | EDF | 15 – 20% improvement in tasks execution time by incorporating placement algorithm |
| [12] | × | | × | | × | | × | | × | × | × | × | Frequency of tasks | Reduction of configuration overheads and improvement of total execution time by decreasing task rejection ratio. schedulable bound of these algorithms is less than 100% |
| . [13] | × | | | × | | × | | × | | × | × | × | Probability of task recurrence in future | Increment in probability of configuration reusing and reduction in overall execution time of tasks. |
| | | | | | | | | | | | | EDF- | Designed for pre- |
| [14] | × | | | × | | × | | × | | × | × | × | Next Fit (EDF-NF) and Merge Server Distribute Load (MSDL) | emptive periodic tasks where MSDL constructs a set of servers for parallel execution and the servers are then scheduled based on EDF-NF for sequential execution |
| [9] | × | | | × | | × | × | | | | × | | Reduction in Execution time. Distance to next call | Introduce a new parameter called Time-Improvement (combination of cost functions 1 and 2 ) and It accelerates task execution from 4% up to 20% |
| [6] | | × | | × | | × | × | | × | | | × | Reserved time of task | It is integration of task allocation, placement and task migration modules. Minimizes task rejection ratio and increase resource utilization |
| [15] | × | × | | × | | | × | × | | | × | × | Level of task graph & task dependencies | Dynamic Scheduling outperforms the Static Scheduling when both task system complexity and degree of dynamism increases. |
| [16] | × | | | × | | | × | × | | | × | × | Distance to the next call and frequency of calls in future | Compiler provides viable information to scheduler in the form of Configuration Call Graph (CCG) |
| . [18] | | × | | × | | | × | × | | | × | × | 1. AET difference b/w SW and HW 2. Required DRL area and memory size of tasks | Proposed HW/SW partitioned algorithms for Shared Memory and Local memory target architectures and Execution time is minimized by parallelizing task execution on DRL architectures. |
| . [17] | × | | | × | | | × | × | × | | | × | Data dependency b/w tasks and hardware area | Handles Communication constraints to utilize the free area b/w subsequent invocations of periodic tasks |
| . [30] | | × | | × | | | × | × | | | × | | Task waiting time | The partitioning algorithm migrate the tasks form HW task to SW task and vice versa based on waiting time of tasks |
| [3] | × | | | × | | | × | × | | | × | | EFT | 8 – 38% improvement in execution time |

| [7] [31] | × |  | × |  | × | × |  | × | × | ALAP | Enhancement to discover area utilized on FPGA by introducing parallelism |

## 7. CONCLUSION AND FUTURE SCOPE

Optimization of real time applications can be done only when HSCS has multifarious resources to support parallel processing where the scheduling algorithms play crucial role in distribution of tasks to the HSCS resources. In this paper, we have demonstrated various High Speed Computing systems which support runtime requirements of applications and also prepared a summary chart for existing scheduling methodologies. The homogeneous computing systems provide parallel processing to the applications at the expense of number of resources, the heterogeneous computing systems support distributed application with the expense of communication between resources, the reconfigurable systems brings dynamic reconfiguration in run time to the application at the expense of soft core processor efficiency and finally the HRCS provides optimal solution for computing real time application by integrating both soft core and hardcore processor as computing elements. The summary chart clearly states that the dynamic scheduling methodologies with multitask are effective in speedup real time application on HSCS but the scheduling model could run always on soft core processors which degrades the efficiency of scheduling model in runtime. So, there is a demand for researchers to develop scheduling model which could run on hard core processor which enhances the efficiency (speed) of scheduler in runtime.

## AUTHORS

*Mahendra Vucha* received his B. Tech in Electronics & Comm. Engineering from JNTU, Hyderabad in 2007 and M. Tech degree in VLSI and Embedded System Design from MANIT, Bhopal in 2009. He is currently working for his PhD degree at MANIT, Bhopal and also working as Asst. Prof in Christ University, Dept. of Electronics and Communication Engineering, Bangalore (K.A), India. His areas of interest are Hardware Software Co-Design, Analogy Circuit design, Digital System Design and Embedded System Design.

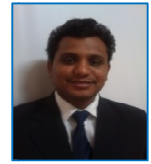

*Arvind Rajawat* received his B. Tech in Electronics & Communication Engineering from Govt. Engineering College in 1989, M. Tech degree in Computer Science Engineering from SGSITS, Indore in 1991 and Ph. D degree from MANIT, Bhopal. He is currently working as Professor in Dept. Electronics and Communication, MANIT, Bhopal (M.P), India. His areas of interest are Hardware Software Co-Design, Embedded System Design and Digital System Design.

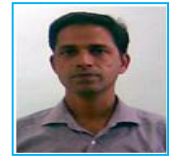